\documentclass[a4paper,10pt]{article}
\pdfoutput=1
\usepackage[utf8]{inputenc}
\usepackage[backend=bibtex,sorting=none]{biblatex}
\usepackage{acronym}
\usepackage{graphicx}
\usepackage{siunitx}
\usepackage{isotope}
\usepackage{hyperref}
\usepackage{authblk}
\usepackage{subfig}

\title{ROAn, a ROOT based Analysis Framework}
\author[1,2,3]{Thomas Lauf}
\author[1,2]{Robert Andritschke\thanks{Corresponding Author, randrits@mpe.mpg.de}}
\affil[1]{Max Planck Institute for Extraterrestrial Physics}
\affil[2]{MPI Halbleiterlabor}
\affil[3]{now at TNG Technology Consulting GmbH}
\date{\today}

\hypersetup{colorlinks=true} 
\begin{document}

\acrodef{depfet}[DePFET]{\textit{Depleted P-channel Field Effect Transistor}}
\acrodef{hll}[HLL]{\textit{MPI Halbleiterlabor}}
\acrodef{mpe}[MPE]{\textit{Max Planck Institute for Extraterrestrial Physics}}
\acrodef{mpp}[MPP]{\textit{Max Planck Institute for Physics}}
\acrodef{roan}[ROAn]{\textit{ROOT based Offline and Online Analysis}}
\acrodef{ccd}[CCD]{\textit{Charge-Coupled Device}}
\acrodef{sdd}[SDD]{\textit{Silicon Drift Detector}}
\acrodef{adu}[ADU]{\textit{Analog-to-Digital Unit}}
\acrodef{gui}[GUI]{\textit{Graphical User Interface}}

\maketitle

\begin{abstract}
The \ac{roan} framework was developed to perform data analysis on data from \ac{depfet} detectors, a type of active pixel sensors developed at the \ac{hll}. \ac{roan} is highly flexible and extensible, thanks to ROOT's features like run-time type information and reflection. \ac{roan} provides an analysis program which allows to perform configurable step-by-step analysis on arbitrary data, an associated suite of algorithms focused on \ac{depfet} data analysis, and a viewer program for displaying and processing online or offline detector data streams. 

The analysis program encapsulates the applied algorithms in objects called steps which produce analysis results. The dependency between results and thus the order of calculation is resolved automatically by the program. To optimize algorithms for studying detector effects, analysis parameters are often changed. Such changes of input parameters are detected in subsequent analysis runs and only the necessary recalculations are triggered. This saves time and simultaneously keeps the results consistent.

The viewer program offers a configurable \ac{gui} and process chain, which allows the user to adapt the program to different tasks such as offline viewing of file data, online monitoring of running detector systems, or performing online data analysis (histogramming, calibration, etc\@.).

Because of its modular design, \ac{roan} can be extended easily, e.g.\ be adapted to new detector types and analysis processes.

\end{abstract}

\acresetall

\section{History}
The \ac{depfet} detectors are the most advanced active pixel detectors produced -- among other detector types such as \acp{sdd} and \acp{ccd} -- at the \ac{hll}, formerly a joint facility by the \ac{mpe} and the \ac{mpp}. 

The \ac{depfet} was invented by Lutz and Kemmer in 1987 \cite{Lutz1987} and a first working prototype was presented in 1990 \cite{Kemmer1990}. Data from the first devices were analysed with the analysis tool \textit{HLLSAS} which was originally written for analysis of \ac{ccd} data. Being a pixelated detector, the \ac{depfet} shares properties of a \ac{ccd}, so the basic data analysis is quite similar. However, when studying detector effects the differences become more dominant and made adaptions necessary. Unfortunately these were hard to install into \textit{HLLSAS}\@. This gave rise for the development of a new data analysis tool. This new tool ought to be modular and extensible such that it can adapt to new devices and analysis needs. The result was the \ac{roan} framework which is presented below.

\section{\acs{roan} in a nutshell}
The \acf{roan} framework offers two programs, \texttt{Analysis} and \texttt{FrameViewer}. The former performs a step-by-step data analysis, the latter is used to browse frame data stored in various formats, as well as to perform online monitoring. Both programs are described in sections~\ref{S:Analysis} and~\ref{S:FrameViewer} below.

\begin{figure}[tb]
   \centering
   \includegraphics[width=0.95\textwidth]{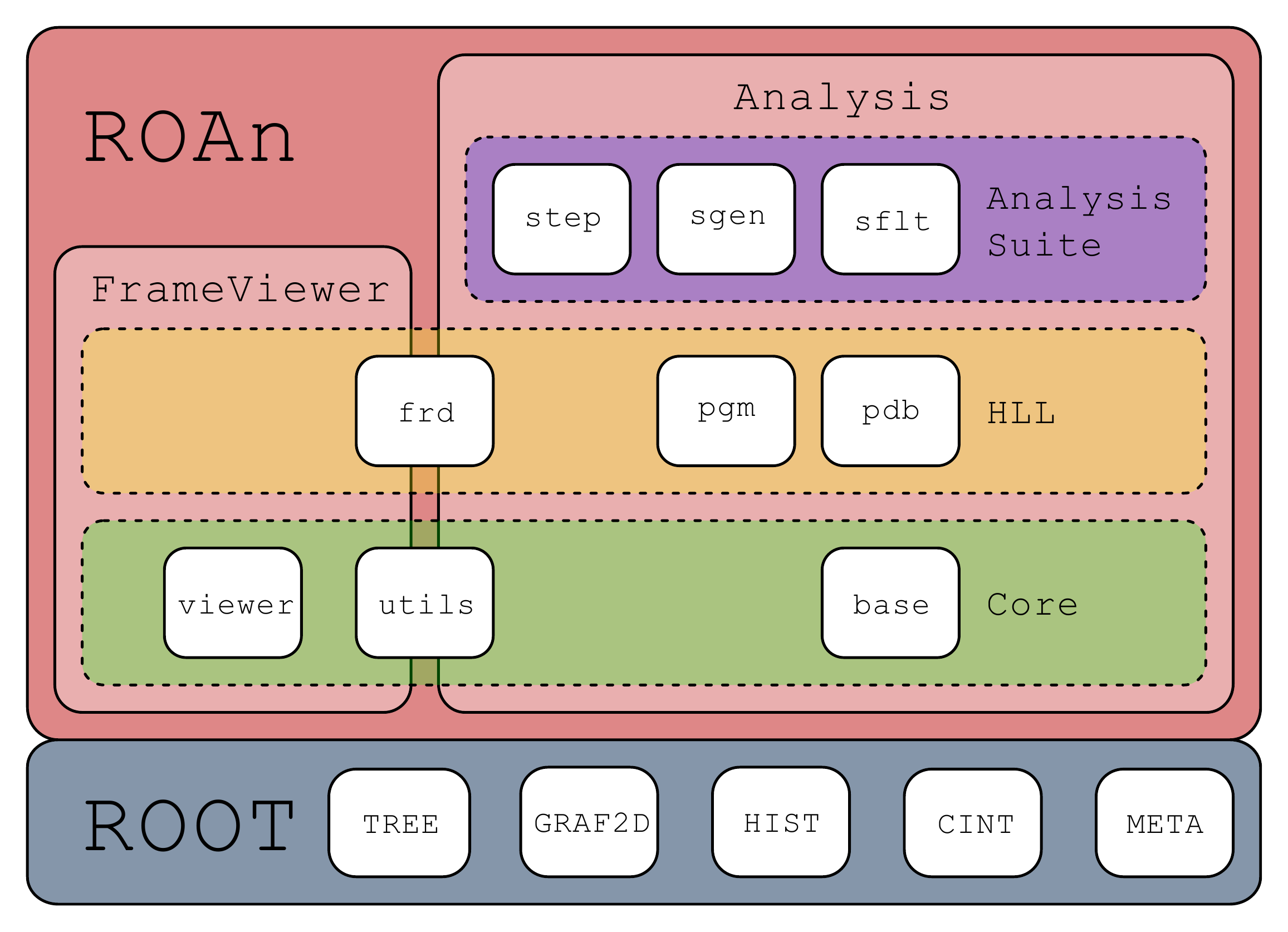} 
   \caption{The architecture of \acs{roan}. The framework is divided into several modules which can be grouped into layers of rising complexity. The \texttt{Core} layer modules provide general functionality, the \texttt{HLL} layer modules generic features such as access to the in-house \textit{framesfile} data format, and the \texttt{Analysis Suite} layer modules contain everything associated with detector-specific data analysis. This modular structure makes the framework highly flexible and easy to extend.}
   \label{F:architecture}
\end{figure}

The \texttt{Analysis} tool is not limited to \ac{depfet} data but can be used to any data where a step-wise approach as described below is feasible. Thanks to the modular structure of the \ac{roan} tool, this can be easily accomplished by using and extending the available interfaces.

The same holds for the \texttt{FrameViewer} program. Currently mainly oriented on the displaying and processing of \ac{depfet} data, its modular and flexible structure lets the user adapt it easily to new tasks.

Figure~\ref{F:architecture} shows an overview of the architecture of \ac{roan}. The \ac{roan} framework makes heavy use of the ROOT framework, especially its tree structure for storing event data, IO, graphics routines, and its C-interpreter \textit{CINT}\@. The use of the latter makes the \ac{roan} tools (especially \texttt{Analysis}) scriptable and enables users to develop new features quickly by writing ROOT macros, which can be dynamically loaded by the interpreter \cite{Brun1997}.

The \ac{roan} framework is divided into several modules which are used by the two applications mentioned above. The modules themselves can be grouped into \textit{layers} which more or less depend on each other.

The bottom layer, named \texttt{Core} in figure~\ref{F:architecture}, comprises the basic modules. These contain mostly abstract interfaces, general utility functions, and the central engine to run the respective application (details can be found below in sections~\ref{S:Analysis} and~\ref{S:FrameViewer}). The implementation of the interfaces is then provided by other modules.

The middle layer, named \texttt{HLL}, groups modules which are more specific than the core modules but more general than the top layer modules. For example, the frame reader module \texttt{frd} encapsulates the access to frames files in which the \ac{depfet} detector data is stored. However, this in-house data format is not limited to this type of detector and can thus be used by analysis suites for different detector types as well.

The top layer, named \texttt{Analysis Suite}, contains the concrete implementation of the analysis algorithms. The modules shown in figure~\ref{F:architecture} are those for \ac{depfet} data analysis. For another type of detector, a different analysis suite can be implemented which still uses the lower layers (e.g.\ for \ac{ccd} detectors, which also store their data in the frames file format, so one can reuse the frame reader module).

The core modules are simple and universal, detector specific details are encapsulated in higher level modules. This modular structure of \ac{roan} makes it easy to adapt the framework to new tasks. 

\section{The \texttt{Analysis} tool}\label{S:Analysis}
\begin{figure} 
   \centering
   \includegraphics[width=0.95\textwidth]{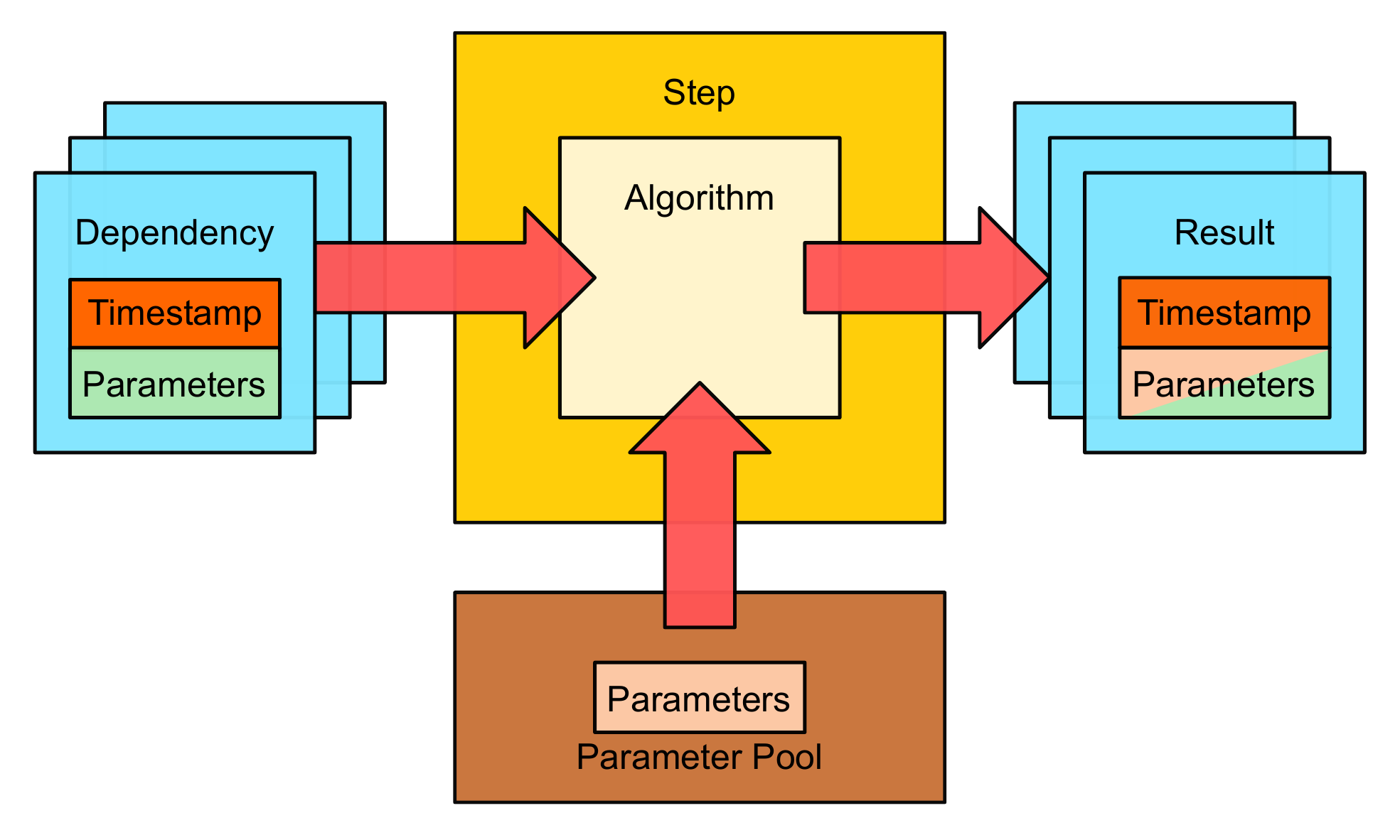} 
   \caption{Within the \texttt{Analysis} program, the algorithms are encapsulated into modules called \textit{steps}. Each step produces one or more results, for which the algorithm uses a set of parameters and requires zero or more results from other steps (so-called \textit{dependencies}). Each result is stored with a timestamp and the parameters used for its calculation. This system enables up-to-date checks for analysis results and can trigger automatic recalculation, thus easing the analysis task for the user.}
   \label{F:analysisstep}
\end{figure}

The \texttt{Analysis} tool performs a step-by-step data analysis in which the algorithms applied to the data are encapsulated in objects called `steps' and executed by a central unit. As shown in figure~\ref{F:analysisstep}, each step produces a set of one or more results and requires a set of zero or more results. The latter ones are thus called \textit{dependencies}. Additionally, parameters are used to fine tune and to provide additional information for the applied algorithm. Each result is stored along with a timestamp and the parameter set used to calculate it. The parameter set of a result is not only the set of parameters used for the algorithm which calculated it, but also contains the union of all parameter sets of its dependencies. 

The \texttt{Analysis} tool can query the steps about their dependencies, the necessary parameters, and keeps track about each result and its timestamp and parameter set. When a user requests the calculation of a specific result, the \texttt{Analysis} tool can resolve the required dependencies and, if necessary, trigger their calculation as well. This is done iteratively for every dependency. 

As shown above, each analysis result is the product of its parameters and its dependencies. A change in either of them therefore voids this specific analysis result and makes its re-calculation necessary. Such a change is introduced for example when modifying thresholds to study detector effects or changing energy ranges for a spectrum.

Whether recalculation is necessary or not is resolved by comparing both the timestamps and parameter sets of the step's results and dependencies. As the dependencies have to be calculated before the step's algorithm can be executed, the timestamps of the dependencies have to be smaller than the timestamps of the results. A dependency with a larger timestamp than a result means that it has been recalculated and is thus newer. 

The ability to resolve dependencies between results, checking whether a result is up-to-date, and to automatically trigger necessary re-calculations, gives the \texttt{Analysis} tool a behaviour similar to the build management tool ``make''. This way the user is relieved from the task of checking result consistency, unnecessary recalculations, and can concentrate on data analysis.

\section{Standard processing of \acs{depfet} data}
The behaviour described in the previous section is implemented in the module \texttt{base} shown in figure~\ref{F:architecture}. This section describes the analysis suite for \ac{depfet} detectors which is implemented in the \texttt{Analysis Suite} layer.

\subsection{\acs{depfet} detector readout}
For the purpose of this paper, it is enough to know that the \ac{depfet} detector matrix is read out column-parallel row by row. One pass of this \textit{rolling shutter} through the matrix results in a \textit{frame} which contains the readout data of all pixels. Further details about the \ac{depfet} detector readout can be found in \cite{Treis2006b,Porro2008,Porro2009}.

\begin{figure}[hbt]
   \centering
   \includegraphics[width=0.45\textwidth]{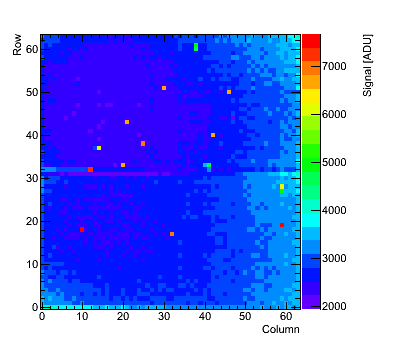} 
   \caption{A typical detector frame from a detector with two hemispheres. The upper and lower half of the detector are each read out with a dedicated readout chip. In the image single photon signals can be seen. The extraction and analysis of these signals is one task executed by the \ac{depfet} analysis suite.}
   \label{F:frame}
\end{figure}

Figure~\ref{F:frame} shows a typical image of a single frame from the detector data stream. The purpose of data analysis is to process this data stream and discriminate the photon signals from pixel noise, calibrate the data, scan for detector effects and apply corrections for them, evaluate the data, and finally report the characteristic numbers necessary for further developing and tuning of the examined detector system.

\subsection{Offset and noise}
The pixel offset is defined as the mean of the detector data when no photons hit the detector. The pixel noise is defined as the standard deviation of the same detector data after subtracting the pixel offset and common mode noise (see section~\ref{S:commonmode}). The simplest way to calculate them is from a set of frames without illumination.

\begin{figure}%
\centering 
\subfloat[Offset Map]{%
\includegraphics[width=0.45\textwidth]{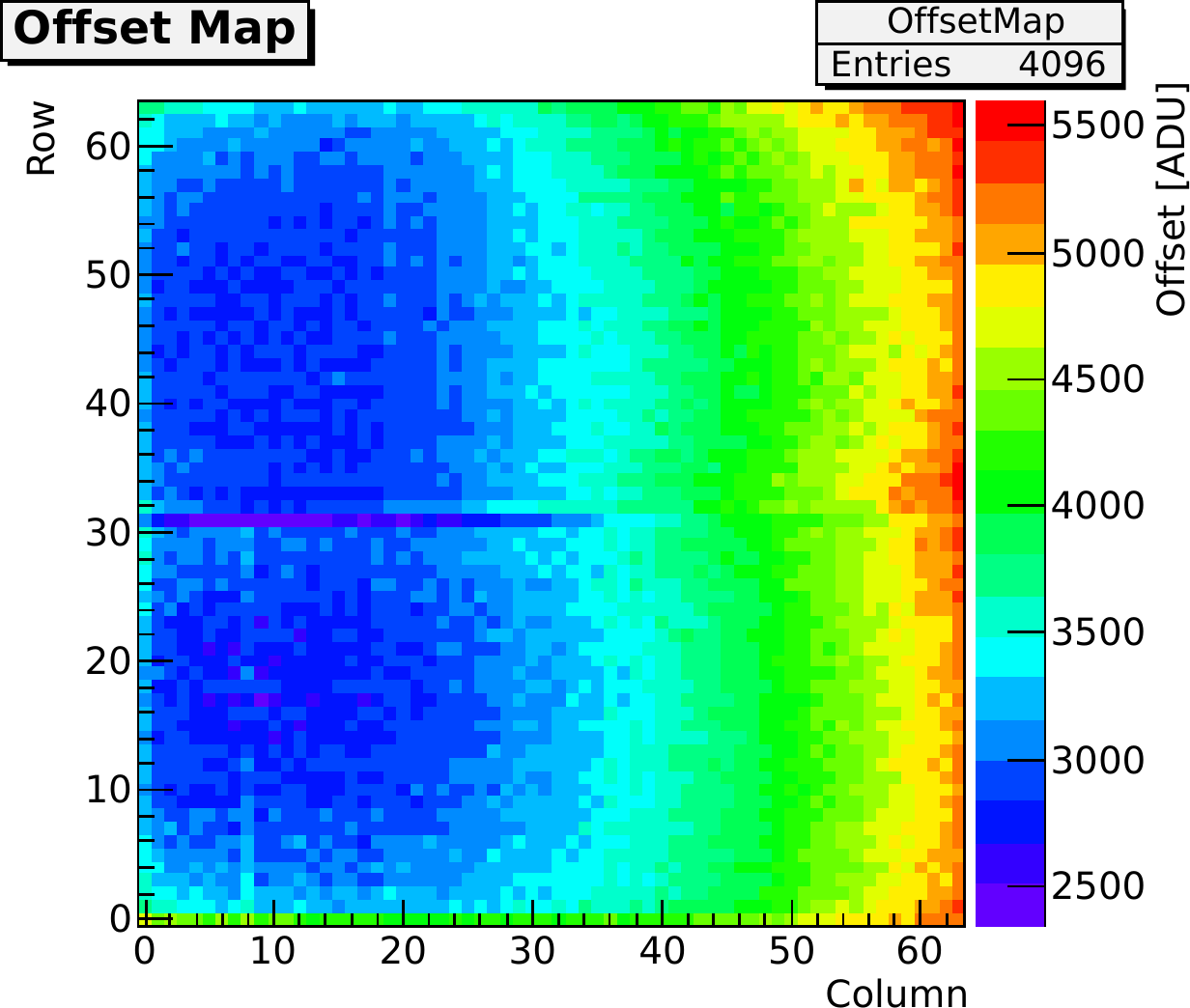}%
}\qquad 
\subfloat[Noise Map]{%
\includegraphics[width=0.45\textwidth]{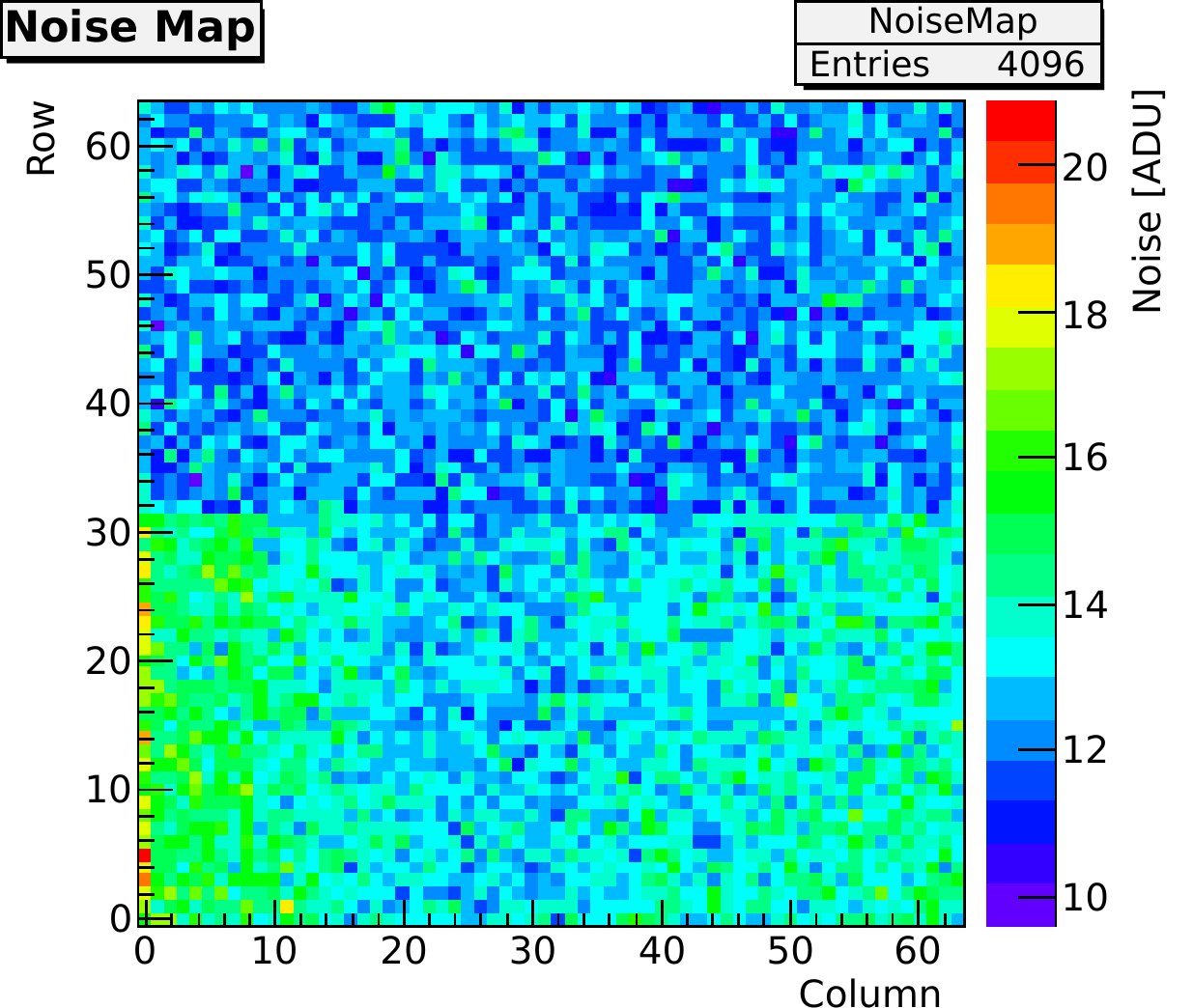}%
}\\ 
\caption{Examples for offset map and noise map. Offset and noise are determined for each pixel in the matrix. They are defined as the average and standard deviation of the signals coming from an unilluminated pixel, respectively.}
\label{F:offsetnoise} 
\end{figure}

Figure~\ref{F:offsetnoise} shows the offset map and the noise map for the same matrix from which the frame image shown in figure~\ref{F:frame} was taken. The detector is divided into two hemispheres which are read out by a dedicated readout chip each.

\subsection{Common mode correction}\label{S:commonmode}
The \ac{depfet} matrix columns are read out in parallel. Any noise contribution that affects all channels in common causes a deviation on a line-by-line basis. An example for such a contribution are fluctuations on supply lines. After offset subtraction this \textit{common-mode noise} is subtracted by calculating and subtracting the median of the current row from each pixel. 

Of course, this algorithm only works if the number of hits in each row is much less than the number of pixels. In cases higher occupancy, other methods have to be deployed.

\subsection{Hit filtering and pattern clustering}
\begin{figure}[h]%
\centering 
\subfloat[Frame after offset subtraction]{\label{F:offsetsub}%
\includegraphics[width=0.45\textwidth]{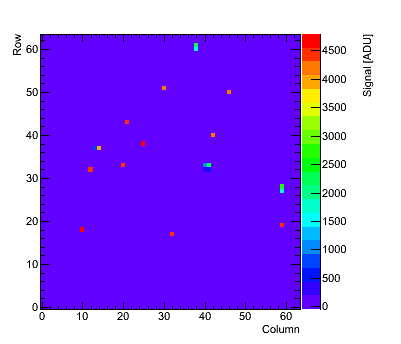}%
}\qquad 
\subfloat[Frame after threshold application]{\label{F:thresapp}%
\includegraphics[width=0.45\textwidth]{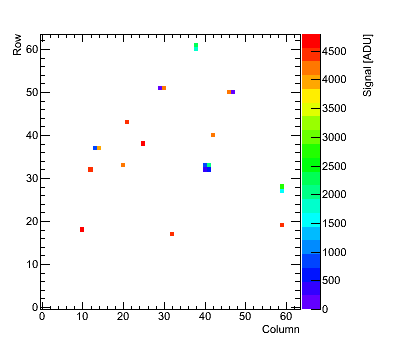}%
}\\ 
\caption[Frame processing]{The frame from figure~\ref{F:frame} after subtraction of the offset and common-mode noise {\subref{F:offsetsub}} and threshold application \subref{F:thresapp}. Hits shown in {\subref{F:thresapp}} which share a common border are then combined to patterns and stored in a list upon which further data analysis takes place.} 
\label{F:hitfiltering} 
\end{figure}

A frame contains the signals of a single matrix readout. To discriminate signals caused by photon events from pixel noise, the offset and common-mode noise is subtracted, and a pixel-wise threshold is applied. Signals above this threshold are considered as a hit caused by a photon, signals below as pixel noise. The results of these steps are shown in figure~\ref{F:hitfiltering}.

Usually, the threshold is set to five times the standard deviation of the respective pixel noise. Finally, the hits are clustered to patterns by combining those which share a common pixel border. This pattern reconstruction helps to increase the peak to background ration by reconstructing charge splitting between pixels.

\subsection{Calibration}
The \ac{depfet} signals are stored in \acp{adu}. Using an illumination source with a distinctive peak one can assign its \ac{adu} value to the energy of the peak and thus calibrate the data. Usually this is done by illuminating the detector by a source with a known energy, like \isotope[55]{Fe} which has a prominent peak at \SI{6}{\kilo\electronvolt}.

For the \ac{depfet} detector, where each pixel represents a readout node, a calibration value for each pixel has to be found. This means that each pixel has to gather enough statistic such that the \ac{adu} value of the corresponding peak can be determined. This can be quite challenging depending on pixel size, detector type, and application.

\subsection{Spectra generation}
\begin{figure}[h]%
\centering 
\subfloat[Detector energy resolution]{\label{F:energyres}%
\includegraphics[width=0.8\textwidth]{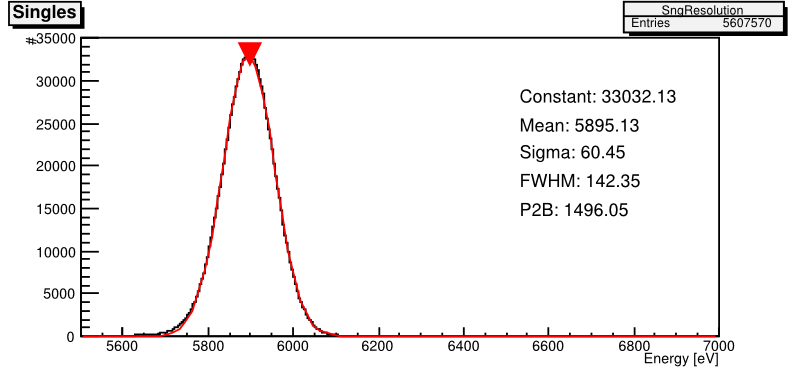}%
}\\ 
\subfloat[Detector hitmap]{\label{F:hitmap}%
\includegraphics[width=0.4\textwidth]{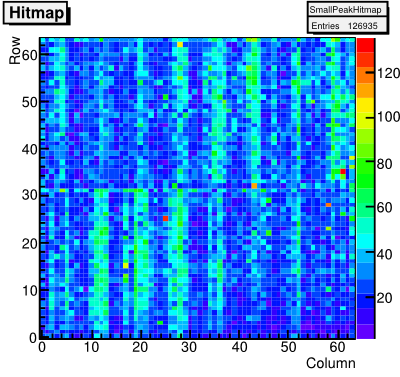}%
}\qquad
\subfloat[Pattern statistics]{\label{F:patternstats}%
\includegraphics[width=0.4\textwidth]{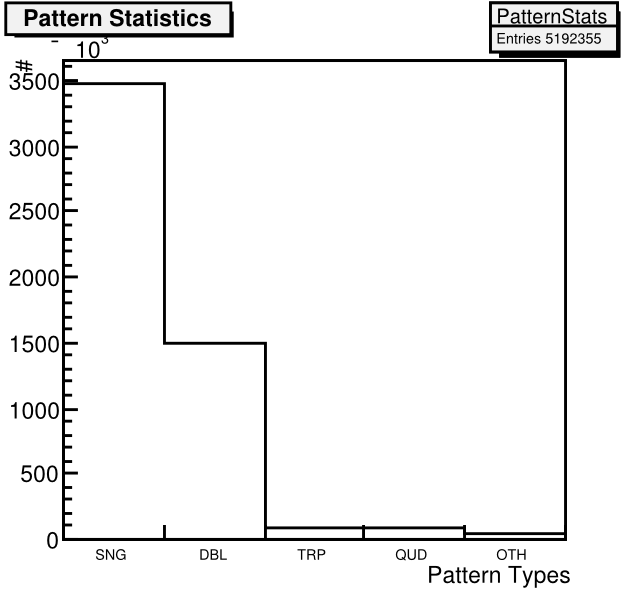}%
}
\caption[Analysis results]{A selection of analysis results. The pattern data extracted from the frame data set is histogrammed to retrieve spectra such as the one shown in \subref{F:energyres} from which the detector energy resolution can be determined. Other result types such as hitmaps showing the spatial distribution of patterns \subref{F:hitmap}, as well as statistical histograms \subref{F:patternstats} can be produced.} 
\label{F:specgen} 
\end{figure}

The \ac{depfet} data can be histogrammed to produce spectra which illustrate different qualities of the detector. These can be used further to determine detector parameters such as the energy resolution or the peak-to-background ratio as shown in figure~\ref{F:specgen}. 

Further the data can be used to investigate pattern distribution by generating hitmaps, applying energy cuts, and much more. Also here a generic flexible system is available which makes creation of new types of spectra easy. For more details on this, the reader is referred to the \ac{roan} Users Guide.

\section{The \texttt{FrameViewer}}\label{S:FrameViewer}
Originally written to view frames files, the \texttt{FrameViewer} has developed into a flexible, multi-purpose viewer, which is now also used for detector online monitoring.

Again, flexibility is achieved through modularity here. The individual processing steps -- frame readout, offset subtraction, threshold application, pattern clustering, and so on -- are encapsulated into small units called \textit{frame factories}. \ac{gui} tasks, such as receiving user input and displaying results, are encapsulated into \textit{viewer modules}.

In contrast to the steps described in section~\ref{S:Analysis}, which are used by the \texttt{Analysis} tool and perform their calculation on the whole frame data set, frame factories perform one processing step for each frame in the data set.

\begin{figure}[h]%
\centering 
\includegraphics[width=0.95\textwidth]{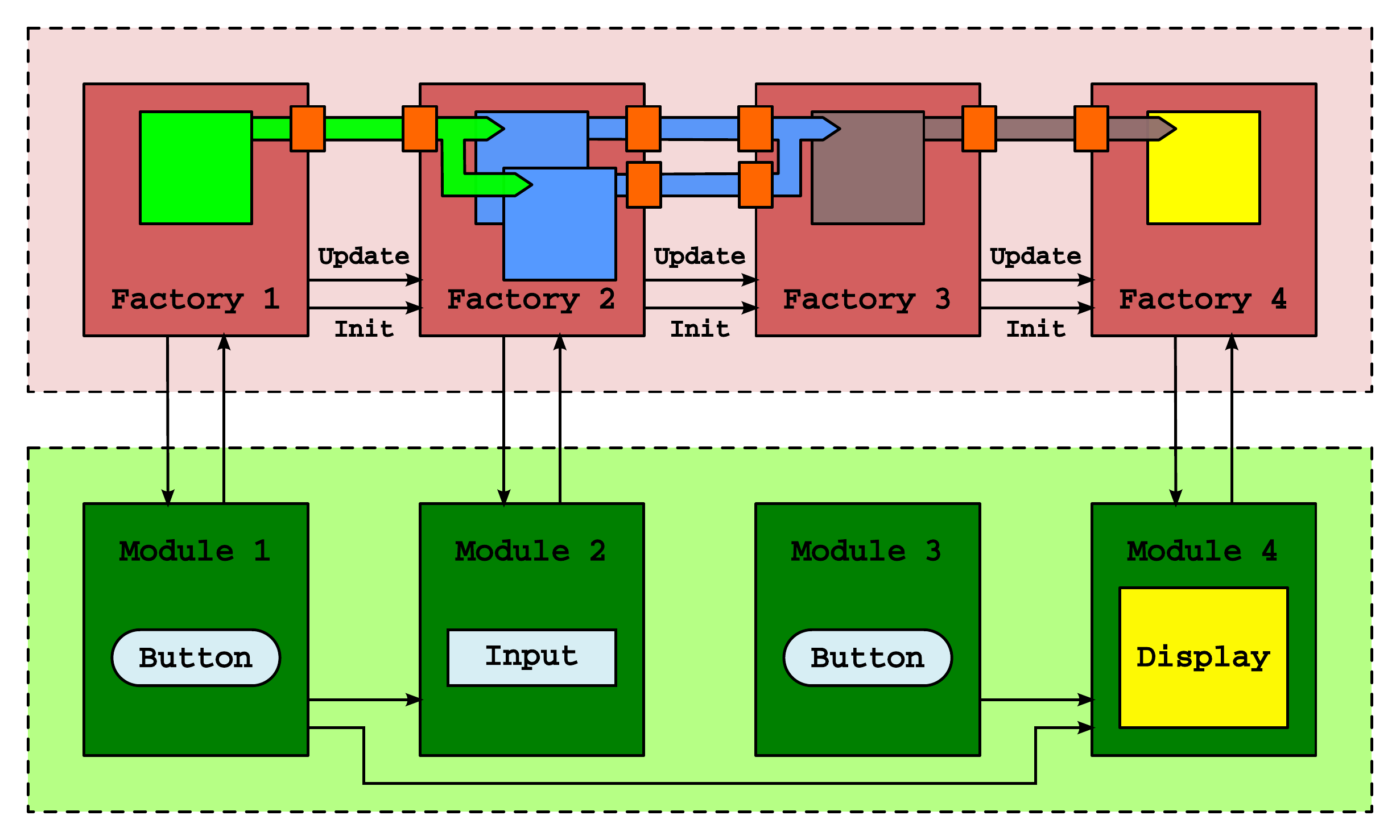}%
\caption{A simple example for a \texttt{FrameViewer} configuration. Above a frame processing chain consisting of four frame factories, below a \ac{gui} with four viewer modules is shown. Frame factories and viewer modules use signals to exchange information among themselves and with each other (black arrows).} 
\label{F:FrameViewer} 
\end{figure} 

The frame factories have internal buffers in which they store their processing results. Other frame factories can access these buffers and use the data as input. The factories have a common, generic interface which makes it possible to chain them. This way, processing chains for arbitrary tasks can be set up.

The frame factories and the viewer modules exchange signals with each other via the \textit{signal-slot-mechanism}\footnote{https://en.wikipedia.org/wiki/Signals\_and\_slots} known from the Qt-Framework which is also implemented in ROOT.

A simplified description of the functionality of the \texttt{FrameViewer} program is now given. For further details  the reader is referred to the \ac{roan} Users Guide. Figure~\ref{F:FrameViewer} shows a graphical representation of the \texttt{FrameViewer}'s architecture. The above row shows a chain of four frame factories, below a \ac{gui} with four viewer modules is sketched. 

When the frame processing chain is set up, the first factory in the chain initializes itself and emits a signal to the next factory (arrow named ``Init'' in figure~\ref{F:FrameViewer}). This factory then initializes itself and emits this signal as well. This way the whole chain is initialized. The initialization process is repeated whenever a vital parameter, for example dimensions of the internal buffer, changes.

The frame processing starts in factory~1. It fills its buffer with data and then signals factory~2 that new data is available (arrow named ``Update'' in figure~\ref{F:FrameViewer}). Factory~2 accesses the data in factory~1 and starts its own processing, thus filling its two buffers. After completing its processing factory~2 signals factory~3, which processes the data and signals factory~4.

When factory~4 is done with processing the signal is returned to factory~3, which returns it to factory~2, which returns it to factory~1 (the signal return is not explicitly shown in figure~\ref{F:FrameViewer}). When factory~1 receives back the signal it fills its buffer with new data and restarts the processing chain by signalling factory~2 again. So for each new frame that is filled into the buffer of factory~1 it starts the processing chain which places the final result in the buffer of the last frame factory (which for example displays it on screen).

Everything connected to a \ac{gui} task in the \texttt{FrameViewer} tool is encapsulated in a \textit{viewer module}. A viewer module can contain a button such as module~1 and module~3 in figure~\ref{F:FrameViewer}, or an input field such as module~2, or a display such as module~4. 

Modules exchange signals with the frame factories but also among themselves to pass information about initialization state, status change, and so on. In figure~\ref{F:FrameViewer} module~1 sends signals to module~2 and module~4, module~4 also receives a signal from module~3.

As shown above, the frame factories encapsulate data processing steps, the viewer modules encapsulate \ac{gui} tasks. Both communicate with each other via signals. Some signals are predefined such as those for new data available and initialization, but they are not limited to these.

The frame factories write their processing results into internal buffers which can be accessed by other frame factories. The interface is realized via so-called ports, which hold type information and the dimension of the buffer. As long as the type and the dimension of the input and output ports of the frame factories match they can be connected.

\begin{figure}[]%
\centering 
\subfloat[]{\label{F:gui1}%
\includegraphics[width=0.4\textwidth]{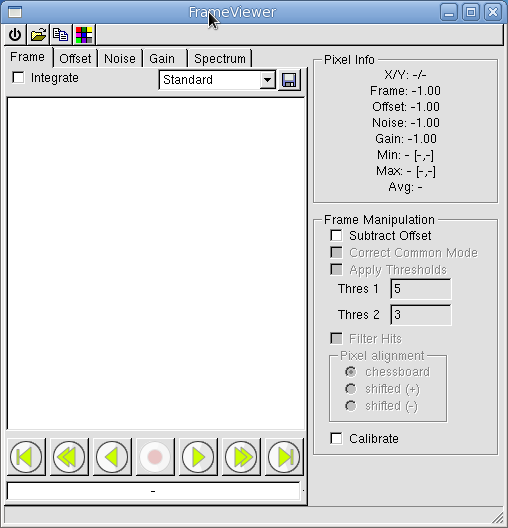}%
}\qquad
\subfloat[]{\label{F:gui2}%
\includegraphics[width=0.4\textwidth]{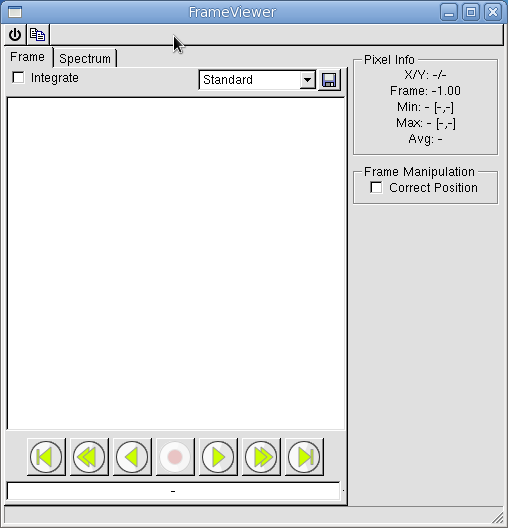}%
}
\caption[FrameViewer GUI examples]{Two examples of different \acsp{gui} of the \texttt{FrameViewer} program. \subref{F:gui1} shows a standard \acs{gui} for viewing offline data, \subref{F:gui2} a \acs{gui} used for online monitoring. The latter has less features and uses other viewer modules than the former. However, there are also viewer modules used by both \acsp{gui} such as the large display area. The \texttt{FrameViewer} \acs{gui} is defined by a configuration text file and thus can be adapted to the specific task without the need to recompile the program.} 
\label{F:FrameViewerGUI} 
\end{figure} 

The setup of the factory chain and the individual viewer modules, as well as the interconnections between frame factories and viewer modules, is described in a configuration file which is plain ASCII text. This configuration file is parsed at program start-up and the frame factories and viewer modules are created, set up, and configured according to the instructions found. Two examples for such \acp{gui} are shown in figure~\ref{F:FrameViewerGUI}.

This way, the \texttt{FrameViewer} is loading only the needed frame factories and viewer modules which makes it ultimately flexible and adaptable, but also gives it a clearly arranged \ac{gui} containing only the viewer modules necessary for the task at hand.

\section{Outlook}
The \ac{roan} framework development has started in \num{2006}. Now the framework has reached an advanced state and has been used applied and used for ASIC qualifications~\cite{Bombelli2009, Bombelli2010} and detector qualifications in projects such as XEUS~\cite{Lauf2009}, Simbol\hbox{-}X~\cite{Lechner2008}, IXO~\cite{Meuris2010, Stefanescu2010}, and BepiColombo~\cite{Porro2009, Majewski2012}. A standard set of analysis steps for \ac{depfet} data analysis is available and has been adapted continuously to the latest developments of \ac{depfet} detectors.

Further developments of the \ac{roan} framework take place on each of the levels shown in figure~\ref{F:architecture}: The central core of the analysis framework, the \ac{hll} specific modules, and the analysis step suite itself. This comprises improvements and standardisation of the interfaces, new modules for future \ac{depfet} matrices, as well as implementation of new algorithms to study future detector effects. Other future goals are improvements such as meta-analysis (e.g.\ comparison of results from different runs), integration of external databases, or parallel computing to accelerate the analysis itself.

\ac{roan} is by design not limited to \ac{depfet} data analysis. Technically it can be applied to any type of data analysis where a step based approach is feasible, e.g. \ac{ccd} or \ac{sdd} detector data. This flexibility makes \ac{roan} a valuable data analysis tool for further developments at \ac{hll} and \ac{mpe}.

\printbibliography

\end{document}